\documentclass[prd,aps,floats,amssymb,preprint,preprintnumbers,nofootinbib]{revtex4}

\usepackage{epsfig}
\usepackage{amssymb}
\usepackage{bbm}
\usepackage{color}
\usepackage{ulem}

\usepackage{enumitem}
\usepackage{graphicx}
\usepackage{subfigure}
\usepackage{amsmath}
\usepackage{bbm}
\usepackage{color}

\def\be{\begin{equation}}
\def\ee{\end{equation}}
\def\bea{\begin{eqnarray}}
\def\eea{\end{eqnarray}}

\begin{document}


\title{Cascading Cosmology}
\author{Nishant Agarwal$^1$}
\author{Rachel Bean$^1$}
\author{Justin Khoury$^2$}
\author{Mark Trodden$^2$}

\affiliation{$^1$Department of Astronomy, Cornell University, Ithaca, NY 14853, USA \\
$^2$Center for Particle Cosmology, Department of Physics and Astronomy, University of Pennsylvania,
Philadelphia PA 19104, USA}

\date{\today}

\begin{abstract}
We develop a fully covariant, well-posed $5D$ effective action for the $6D$ cascading gravity brane-world model, and use this to study cosmological solutions. We obtain this effective action through the $6D$ decoupling limit, in which an additional scalar degree mode, $\pi$, called the brane-bending mode, determines the bulk-brane gravitational interaction. The $5D$ action obtained this way inherits from the sixth dimension an extra $\pi$ self-interaction kinetic term. We compute appropriate boundary terms, to supplement the $5D$ action, and hence derive fully covariant junction conditions and the $5D$ Einstein field equations. Using these, we derive the cosmological evolution induced on a 3-brane moving in a static bulk.
We study the strong- and weak-coupling regimes analytically in this static ansatz, and perform a complete numerical analysis of our solution. Although the cascading model can generate an accelerating solution in which the $\pi$ field comes to dominate at late times, the presence of a critical singularity prevents the $\pi$ field from dominating entirely. Our results open up the interesting possibility that a more general treatment of degravitation in a time-dependent bulk, or taking into account finite brane-thickness effects, may lead to an accelerating universe without a cosmological constant.
\end{abstract}

\maketitle

\section{Introduction}
\label{intro}

A fundamental conundrum exists as to whether the accelerated expansion of the universe is due to a new form of energy or novel gravitational physics revealing itself at ultra-large scales, extremely low spatial curvatures, and low cosmological densities. Along with studies of different forms of dark energy and modifications to gravity, considerable attention has been paid to the possible role played by higher-dimensional theories, in which our four-dimensional world is considered to be a surface (a ``brane") embedded in a higher-dimensional spacetime (the ``bulk"). In the old Kaluza-Klein picture it was necessary for the extra dimensions to be sufficiently compact (for reviews see, e.g.~\cite{Bailin:1987jd,Overduin:1998pn}). Recent developments, however, are based on the idea that all standard model particles are confined to a $4D$ brane, whereas gravity is free to explore the bulk~\cite{Lukas:1998yy,ArkaniHamed:1998rs,Antoniadis:1998ig}. As such, ``large" extra dimensions are conceivable, giving rise to a much smaller fundamental Planck mass than the effective Planck scale we observe today~\cite{ArkaniHamed:1998rs,Randall:1999ee,Randall:1999vf}. A well-studied example of such a theory is the Dvali-Gabadadze-Porrati (DGP) model \cite{Dvali:2000hr}, in which our observed $4D$ universe is embedded in an infinite fifth dimension. In this picture, the higher-dimensional nature of gravity affects the $4D$ brane through deviations from general relativity on horizon scales, $r\sim cH_0^{-1}$ (where $c$ is the speed of light and $H_{0}$ is the Hubble constant), that may give rise to the observed accelerated expansion. 

In the DGP model, integrating out the bulk degrees of freedom yields an effective action for the $4D$ fields containing, besides the graviton, an extra scalar degree of freedom, $\pi$, called the brane bending mode~\cite{Luty:2003vm,Nicolis:2004qq,Deffayet:2001uk}. The $\pi$ field contributes to the extrinsic curvature of the boundary and interacts strongly at the energy scale $\Lambda_{5} = M_{5}^{2}/M_{4}$, where $M_{5}$ and $M_{4}$ are the $5D$ and $4D$ Planck masses respectively. In analogy with massive gravity~\cite{ArkaniHamed:2002sp}, there exists a decoupling limit in which the strong interaction scale $\Lambda_{5}$ is held fixed while $M_{4},M_{5} \rightarrow \infty$. All other degrees of freedom (including the graviton and a vector $N_{\mu}$) decouple in this limit. This implies that the dynamics of the scalar field $\pi$ can completely describe all interesting features of the DGP model, including the Vainshtein screening effect~\cite{Deffayet:2001uk} and the self-accelerated cosmological solution~\cite{Deffayet:2000uy}. It has now been established that the branch of solutions that include self-acceleration suffers from ghost-like instabilities~\cite{Nicolis:2004qq,Koyama:2005tx,Gorbunov:2005zk,Charmousis:2006pn,Dvali:2006if,Gregory:2007xy}. On the observational front, DGP cosmology is statistically disfavored in comparison to $\Lambda$CDM~\cite{Rydbeck:2007gy,Fang:2008kc,Lombriser:2009xg} and is significantly discordant with constraints on the curvature of the universe~\cite{Guo:2006ce}. 

Recently, a phenomenological approach to the cosmological constant problem --- {\it degravitation} \cite{Dvali:2002pe,ArkaniHamed:2002fu,Dvali:2007kt} --- has been developed. In degravitation it is postulated that the cosmological constant is indeed responsible for dark energy. The cosmological constant problem that the observed value is at least 120 orders of magnitude smaller than vacuum energy density predicted theoretically, is solved not by making the \textit{vacuum energy density} small, but instead, by having a large cosmological constant whose gravitational effect is suppressed by making \textit{gravity} extremely weak on large scales. The DGP model should, in principle, provide a more fundamental implementation of degravitation. However, the weakening of gravity observed in DGP is insufficient to account for the disparity between the expected and observed values of the cosmological constant. This fact, in addition to the above mentioned problems of the DGP model, have led to the idea of {\it cascading DGP} \cite{deRham:2007xp,deRham:2007rw,deRham:2009wb,Corradini:2007cz,Corradini:2008tu} --- a higher-dimensional generalization of the DGP idea, which is free of divergent propagators and ghost instabilities. In this model one embeds a succession of higher-codimension branes into each other, with energy-momentum confined to the $4D$ brane and gravity living in higher-dimensional space. 
(See~\cite{Kaloper:2007ap} for a related framework.)

The implemention of degravitation within the cascading gravity idea provides an intriguing new theoretical avenue for solving the problem of dark energy. However an important litmus test is whether such models can reproduce a successful cosmological evolution. Studies thus far in this direction have assumed an effective $4D$ cosmology for degravitation by generalizing that for DGP~\cite{Afshordi:2008rd,Khoury:2009tk}. However, to perform a more complete study of cosmology on the brane, it is necessary to integrate out the sixth and fifth dimensions to obtain a $4D$ effective theory. 

In this paper we start from the action for cascading gravity in $6D$ and obtain an effective linearized $5D$ action in the decoupling limit. This gives rise to an extra brane-bending scalar degree of freedom (the $\pi$ field) in the $5D$ action. As a proxy for the complete $6D$ cascading set-up, we propose a $5D$ non-linear and covariant completion of the quadratic action. A similar strategy was used in~\cite{Chow:2009fm}, where an analogous $4D$ covariant action was shown to reproduce much of the phenomenology of the full DGP model. In our case, the resulting action is a $5D$ scalar-tensor theory, describing $5D$ gravity and a scalar $\pi$, coupled to a $4D$ brane. Because of its scalar-tensor nature, the standard Israel junction conditions must be revisited. We derive the appropriate junction conditions across the $4D$ brane using two different techniques. These can then be used in conjunction with the bulk equations to study cosmology on the brane. For concreteness, we consider the cosmology induced on a moving brane in a static bulk geometry. We find analytical solutions in the strong- and weak-coupling regime for the $\pi$ field, and numerically integrate the full equations of motion. Thanks to the Vainshtein screening mechanism, the resulting $4D$ cosmology is consistent with standard big bang expansion history at early times, but deviates from $\Lambda$CDM at late times. We find that $\pi$ contributes to cosmic acceleration at late times, but a singularity in the brane embedding prevents $\pi$ from accounting for all of dark energy.
 
In section~\ref{cascgravity} we outline the $6D$ cascading gravity model we consider, and propose an effective, covariant $5D$ action with a strongly interacting $\pi$ field that encodes the $6D$ physics. In section~\ref{boundary} we derive the appropriate boundary terms necessary in order for our action to have a well-defined variational principle. The resulting bulk equations of motion and brane junction conditions are computed in section~\ref{eoms}. We then turn in section~\ref{cosmology} to the search for cosmological solutions on the brane, by considering its motion in a static bulk. Finally, we draw together our findings and discuss implications in section~\ref{conclusions}. 

A comment on our notation: we denote coordinates in the full six dimensional spacetime by $x^{0},x^{1},x^{2},x^{3},x^{5},x^{6}$. Indices $M,N,...$ run over 0,1,2,3,5 (i.e. the $4+1D$ coordinates), indices $\mu,\nu,...$ run over 0,1,2,3 (i.e. the $3+1D$ coordinates), and indices $i,j,...$ run over $1,2,3$ (i.e. the $3D$ spatial coordinates). We further denote the fifth and sixth dimensional coordinates by $y=x^{5}$ and $z=x^{6}$, where convenient.


\section{A Proxy Theory for Cascading Gravity}
\label{cascgravity}

The DGP model consists of a 3-brane embedded in a flat, empty $4+1$-dimensional bulk. Despite the fact that the extra dimension is infinite in extent, the inverse-square law is nevertheless recovered at short distances on the brane due to an intrinsic, four-dimensional Einstein-Hilbert term in the action
\be
	S_{\rm DGP} = \int_{\rm bulk} {\rm d}^5x\sqrt{-g_5}\frac{M_5^3}{2}R_5 +
\int_{\rm brane} {\rm d}^4x \sqrt{-g_4} \left(\frac{M_4^2}{2}R_4 + {\cal L}_{\rm matter}\right)\,.
\ee
The Newtonian potential on the brane scales as $1/r$ at short distances, as in $4D$ gravity, and asymptotes to $1/r^2$ at large distances, characteristic of $5D$ gravity. The cross-over scale $m_5^{-1}$ between these two behaviors is set by the bulk and brane Planck masses via
\be
	m_5 = \frac{M_5^3}{M_4^2}\,.
\label{m5}
\ee
From the point of view of a brane observer, this force law arises from the exchange of a continuum of massive gravitons, with $m_5$ setting an effective mass scale for gravity on the brane. The DGP model is therefore a close phenomenological cousin of Fierz-Pauli massive gravity. In particular, brane gravitons form massive spin-2 representations with 5 helicity states, with the helicity-0 mode having a small strong-coupling scale,
\be
\Lambda_5 = (m_5^2M_4)^{1/3}\, .
\ee

There are many reasons to consider extending this scenario to higher dimensions:

\begin{itemize}

\item Pragmatically, cosmological observations already place stringent constraints on the DGP model~\cite{Rydbeck:2007gy,Fang:2008kc,Lombriser:2009xg}. In higher dimensions, however, the modifications to the Friedmann equation are expected to be milder, which traces back to the fact that the $4D$ graviton mass term is a more slowly-varying function of momentum. The resulting cosmology is therefore closer to the $\Lambda$CDM expansion history,
thereby allowing a wider range of parameters.

\item Another motivation, as we have already mentioned, is the degravitation idea~\cite{ArkaniHamed:2002fu,Dvali:2007kt} for addressing the cosmological constant problem; namely that gravity acts as a high-pass filter that suppresses the contribution of vacuum energy to the gravitational field. Although the infrared weakening of gravity displayed in the DGP force law is suggestive of a high-pass filter, in practice this weakening is too shallow to ``filter out" vacuum energy. However, the situation is more hopeful in $D>5$ dimensions, where the force law on the brane falls more steeply as $1/r^{D-2}$ at large distances~\cite{Dvali:2007kt}.

\end{itemize}

While seemingly a straightforward task, generalizing the DGP scenario to higher dimensions has proven challenging. To begin with, the simplest constructions are plagued with ghost instabilities around flat space~\cite{Dubovsky:2002jm,Gabadadze:2003ck}. Another technical hurdle is the fact that the $4D$ propagator is divergent and requires careful regularization~\cite{Geroch:1987qn,deRham:2007pz}. Finally, for a static bulk, the geometry for codimension $N> 2$ has a naked singularity at a finite distance away from the brane, for an arbitrarily small tension~\cite{Dvali:2002pe}.

It was recently shown that these pathologies are absent if the 3-brane is embedded in a succession of higher-dimensional DGP branes, each with their own Einstein-Hilbert term. In the $5+1$-dimensional case, for instance, the 3-brane lies on a 4-brane, with action,
\bea
\nonumber
	S_{\rm cascade} & = & \int_{\rm bulk} {\rm d}^6x\sqrt{-g_6}\frac{M_6^4}{2}R_6 + \int_{\rm 4-brane} {\rm d}^5x\sqrt{-g_5}\frac{M_5^3}{2}R_5 \\
	& & + \int_{\rm 3-brane} {\rm d}^4x \sqrt{-g_4} \left(\frac{M_4^2}{2}R_4 + {\cal L}_{\rm matter}\right)\,.
\label{S6}
\eea
As a result, the force law on the 3-brane ``cascades" from $1/r^2$ to $1/r^3$ to $1/r^4$ as one moves increasingly far from a source, with the $4D\rightarrow 5D$ and $5D\rightarrow 6D$ cross-over scales given respectively by $m_5^{-1}$ and $m_6^{-1}$, with
\be
m_6 = \frac{M_6^4}{M_5^3}\,.
\ee
This {\it cascading gravity} setup is free of the aforementioned pathologies: the theory is perturbatively stable provided that the 3-brane is endowed with a sufficiently large tension~\cite{deRham:2007xp,deRham:2007rw}; the $5D$ Einstein-Hilbert term acts as a regulator for the induced propagator on the 3-brane; and, as has been shown explicitly for $D=6,7$, adding tension on the 3-brane results in a completely smooth bulk geometry (except of course for the delta-function singularities at the brane locations) and leaves the 3-brane geometry flat, at least for sufficiently small tension~\cite{deRham:2009wb}.

The next question is, of course, whether the resulting cosmology is consistent with current observations, and, more interestingly, whether it offers distinguishing signatures from $\Lambda$CDM cosmology. Unfortunately, finding analytical solutions is a hopeless task, even in the simplest $6D$ case, as the bulk metric is generally expected to depend on all extra-dimensional coordinates plus time~\cite{Chatillon:2006vw}.

In this paper, we instead study a $5D$ ``proxy" brane-world theory for $6D$ cascading gravity, consisting of a scalar-tensor theory of gravity in the $5D$ bulk. This is obtained by generalizing the well-known decoupling limit of standard DGP~\cite{Luty:2003vm} to the cascading case. The limit we propose is $M_5,M_6\rightarrow \infty$, with the strong-coupling scale
\be
\Lambda_6 = (m_6^4M_5^3)^{1/7}
\ee
kept fixed. In this limit, the action~(\ref{S6}) may be expanded around flat space, and reduces to a local theory on the 4-brane, describing $5D$ weak-field metric perturbations $h_{MN}$ and an interacting scalar field $\pi$. The latter is the helicity-0 mode of massive gravity on the 4-brane, and has a geometrical interpretation as measuring the extrinsic curvature of the 4-brane in the $6D$ spacetime. The resulting action is~\cite{deRham:2007xp}
\begin{eqnarray}
	S_{\rm decouple} & = & \frac{M_{5}^{3}}{2} \int_{\rm bulk} {\rm d}^{5}x \left[ -\frac{1}{2} h^{MN}(\mathcal{E}h)_{MN} + \pi\eta^{MN}(\mathcal{E}h)_{MN} - \frac{27}{16m_6^2} (\partial\pi)^{2} \Box_{5}\pi \right] \nonumber \\
	& & + \int_{\rm brane} {\rm d}^{4}x \left[ -\frac{M_{4}^{2}}{4} h^{\mu\nu}(\mathcal{E}h)_{\mu\nu} + \frac{1}{2} h^{\mu\nu}T_{\mu\nu} \right]\,,
\label{5dcov1}
\end{eqnarray}
where
\bea
	(\mathcal{E}h)_{MN} & = & -\frac{1}{2} (\Box_{5} h_{MN} - \eta_{MN}\Box_{5} h -\partial_M\partial^Kh_{KN} \nonumber \\
	& & \ \ \ \ \ - \ \partial_N\partial^Kh_{MK} +\eta_{MN}\partial^K\partial^Lh_{KL}+\partial_M\partial_N h)\, 
\eea
is the linearized Einstein tensor in $5D$, and $(\mathcal{E}h)_{\mu\nu}$ that in $4D$. To see that only these terms survive in the decoupling limit, introduce canonically-normalized variables $\pi^c = M_5^{3/2}\pi$ and $h^c_{MN} = M_5^{3/2}h_{MN}$, which have the correct mass dimension for scalar fields in $4+1$ dimensions. The quadratic terms in~(\ref{5dcov1}) become independent of $M_5$ under this field redefinition, whereas the cubic term reduces to $(\partial\pi_c)^2\Box_{5}\pi_c/\Lambda_{6}^{7/2}$. All other interactions in~(\ref{S6}) are suppressed by powers of $1/M_5$, $1/M_6$ and therefore drop out in the decoupling limit.

In using~(\ref{5dcov1}) as our starting point, we are motivated by the fact that nearly all of the interesting features of DGP gravity are due to the helicity-0 mode $\pi$ and can be understood at the level of the decoupling theory~\cite{Nicolis:2004qq,Chow:2009fm}. Of course, as it stands~(\ref{5dcov1}) is restricted to weak-field gravity and is therefore of limited use for cosmological solutions. As our ``proxy" brane-world scenario, we propose to complete~(\ref{5dcov1}) into a covariant, non-linear theory of gravity in $5D$ coupled to a 3-brane. By construction, the weak-field limit of our theory will coincide with~(\ref{5dcov1}). A similar approach was followed in~\cite{Chow:2009fm} to mimic the $5D$ DGP scenario with a proxy effective theory in $4D$. Despite being a local theory in $3+1$ dimensions, the resulting cosmology was found to be remarkably similar to that of the full $4+1$-dimensional DGP framework, both in its expansion history and evolution of density perturbations.

Generalizing the strategy of~\cite{Chow:2009fm} to the cascading gravity framework, we are led to propose the following non-linear completion of~(\ref{5dcov1}):
\bea
\nonumber
	S & = & \frac{M_{5}^{3}}{2}\int_{\rm bulk}{  {\rm d}^{5}x\sqrt{-g_{5}}\left[e^{-3\pi/2}R_{5} - \frac{27}{16m_6^2}(\partial\pi)^2 \Box_{5}\pi \right]} \\
	& & + \int_{\rm brane}{{\rm d}^{4}x\sqrt{-g_{4}}\left[ \frac{M_{4}^{2}}{2}R_{4} + \mathcal{L}_{\rm{matter}} \right]}\,.
\label{5dcov}
\eea
It is straightforward to check that this theory indeed reduces to~(\ref{5dcov1}) in the weak-field limit, and therefore agrees with cascading gravity to leading order in $1/M_5$. (This is most easily seen by working again with the rescaled variables $\pi^c$ and $h^c_{MN}$.) The proposed $5D$ completion is by no means unique, since one could consider a host of $M_5$-suppressed operators which would disappear in the weak-field limit. Our hope is that the salient features of cascading cosmology are captured by our $5D$ effective theory, and that the resulting predictions are at least qualitatively robust to generalizations of~(\ref{5dcov}). 

The effective action~(\ref{5dcov}) must be supplemented with suitable boundary terms in order to yield a well-defined variational principle. Other than a Gibbons-Hawking-York-like term, the form of the cubic term in $\pi$ clearly necessitates its own boundary contribution. In the next section, we derive these boundary terms, which will be essential in deriving the junction conditions.


\section{Boundary Terms in the $5D$ Effective Theory}
\label{boundary}

Because of the form of the cubic term, varying~(\ref{5dcov}) with respect to $\pi$ yields contributions on the 3-brane of the form $\sim (\partial\pi)^2{\mathcal L}_{n} \delta\pi$, where ${\mathcal L}_n$ is the Lie derivative with respect to the normal. Such terms cannot be set to zero by the usual Dirichlet boundary condition, $\delta\pi = 0$, and must therefore be canceled by appropriate boundary terms in order that the action be truly stationary and the variational principle be well-posed. Gravity also requires its own boundary contribution, which is a generalization of the well-known Gibbons-Hawking-York term~\cite{York:1972sj,Gibbons:1976ue}. (We should, of course, also include boundary terms at infinity, but we will ignore these since they do not play any role in the junction conditions.)

To derive the boundary terms, it is convenient to work in the Arnowitt, Deser and Misner (ADM) coordinates~\cite{Arnowitt:1962hi}, with $y$ playing the role of a ``time'' variable,
\begin{eqnarray}
	ds^{2}_{(5)} = N^{2}{\rm d}y^{2} + q_{\mu\nu} ({\rm d}x^{\mu} + N^{\mu}{\rm d}y) ({\rm d}x^{\nu} + N^{\nu}{\rm d}y)\,,
\label{ADSmetric}
\end{eqnarray}
where $N$ and $N_\mu$ are the lapse function and the shift vector, respectively. In the ``half-picture'', the bulk extends from $y=0$ to $\infty$,
and the 3-brane is located at $y=0$, with normal vector $n_{M} = (0,0,0,0,N)$.

In ADM coordinates, the $5D$ Einstein-Hilbert term takes the form
\be
S_{\rm gravity} = \frac{M_{5}^{3}}{2} \int_{y\geq 0} {\rm d}^{4}x {\rm d}y \sqrt{-q}N e^{-3\pi/2} \left[ R_{4} + K^{2} - K_{\mu\nu}K^{\mu\nu} + 2\nabla_{M}\left(n^{N}\nabla_{N}n^{M} - n^{M}K \right) \right]\,,
\label{5dgravaction}
\ee
where $K_{\mu\nu}$ is the extrinsic curvature tensor
\be
K_{\mu\nu} \equiv  \frac{1}{2}{\cal L}_{n} q_{\mu\nu} = \frac{1}{2N} (\partial_{y}q_{\mu\nu} - D_{\mu}N_{\nu} - D_{\nu}N_{\mu})\,.
\ee
Here $D_\mu$ is the covariant derivative with respect to the $4D$ induced metric $q_{\mu\nu}$. 
Unlike standard gravity, the ${\cal L}_{n} K$ term in~(\ref{5dgravaction}) is not a total derivative and must be treated with care. Integrating by parts gives
\bea
\nonumber
S_{\rm gravity} 
&=& \frac{M_{5}^{3}}{2} \int_{y\geq 0} {\rm d}^{4}x {\rm d}y \sqrt{-q}N \left[ e^{-3\pi/2} \left( R_{4} + K^{2} - K_{\mu\nu}K^{\mu\nu} -3K{\cal L}_{n}\pi  \right) - 2\Box_{4}e^{-3\pi/2} \right] \\
& & + \ M_{5}^{3} \int_{y = 0^{+}} {\rm d}^{4}x \sqrt{-q} e^{-3\pi/2} K\, ,
\eea
and the last term must therefore be canceled with a Gibbons-Hawking-York boundary term
\be
\Delta S_{\rm GHY} = - M_{5}^{3} \int_{y = 0^{+}} {\rm d}^{4}x \sqrt{-q} e^{-3\pi/2} K\,.
\label{gravbdy}
\ee

Similar considerations for the $\pi$-sector lead us to require adding the boundary term
\be
\Delta S_\pi = -\frac{27}{32}\frac{M_{5}^{3}}{m_{6}^{2}} \int_{y=0^{+}} {\rm d}^{4}x \sqrt{-q} \left( \partial_{\mu}\pi\partial^{\mu}\pi {\cal L}_{n}\pi + \frac{1}{3} \left( {\cal L}_{n}\pi \right)^{3} \right)\,,
\label{pibdy}
\ee
where
\be
{\cal L}_{n}\pi = N^{-1}(\partial_{y} - N^{\mu}\partial_{\mu})\pi\, .
\ee
Note that in the flat space limit this agrees with the $\pi$ boundary term obtained in~\cite{Dyer:2009yg} in the decoupled theory.

Including~(\ref{gravbdy}) and~(\ref{pibdy}), the full $5D$ action is therefore
\begin{eqnarray}
	S & = & \frac{M_{5}^{3}}{2} \int_{\rm bulk} {\rm d}^{5}x \sqrt{-g_{5}} \left[ e^{-3\pi/2}R_{5} - \frac{27}{16m_6^2}(\partial\pi)^2 \Box_{5}\pi \right] \nonumber \\
	& & - \ M_{5}^{3} \int_{\rm brane} {\rm d}^{4}x \sqrt{-q} \left[ e^{-3\pi/2} K + \frac{27}{32m_{6}^{2}} \left( \partial_{\mu}\pi\partial^{\mu}\pi {\cal L}_{n}\pi + \frac{1}{3} \left( {\cal L}_{n}\pi \right)^{3} \right) \right] \nonumber \\
	& & + \int_{\rm brane}{{\rm d}^{4}x\sqrt{-q}\left[ \frac{M_{4}^{2}}{2}R_{4} + \mathcal{L}_{\rm{matter}} \right]}\,.
\label{5dcovcomplete}
\end{eqnarray}
Although we obtained the boundary terms using the ADM formalism, the result is fully covariant and hence holds in any coordinate system. In particular, given the unit normal vector to the brane $n^M$ in a general coordinate system, the Lie derivative is given by ${\cal L}_{n}\pi = n^{M}\partial_{M}\pi$, and the induced metric by $q_{MN} = g_{MN} - n_Mn_N$. One can check that varying this action with respect to the metric and $\pi$ does not yield any normal derivative terms of the form ${\cal L}_{n}\delta q_{\mu\nu}$ and ${\cal L}_{n}\delta\pi$ on the boundary.


\section{Covariant Equations of Motion On and Off the Brane}
\label{eoms}

Our goal now is to derive the bulk equations of motion and brane junction conditions that result from~(\ref{5dcovcomplete}). (See~\cite{Chamblin:1999ya,Maeda:2000wr,Mennim:2000wv,Barcelo:2000js} for earlier work on junction conditions in scalar-tensor brane-world scenarios.) Starting with the bulk, varying~(\ref{5dcovcomplete}) with respect to the metric yields the Einstein equations
\bea
\nonumber
	e^{-3\pi/2} G_{MN} & = & -\frac{27}{16m_6^2}\left[ \partial_{(M}(\partial\pi)^{2}\partial_{N)}\pi  - \frac{1}{2}g_{MN}\partial^{K}(\partial\pi)^{2}\partial_{K}\pi - \partial_{M}\pi\partial_{N}\pi\Box_{5}\pi \right] \\
	& & - \left( g_{MN}\Box_{5} - \nabla_{M}\nabla_{N} \right) e^{-3\pi/2},
\label{bulkein}
\eea
where $G_{MN}$ is the $5D$ Einstein tensor. The second line is typical of scalar-tensor theories and arises from the non-minimal coupling of $\pi$ to gravity. Varying with respect to $\pi$, meanwhile, gives 
\begin{eqnarray}
	(\Box_{5}\pi)^{2} - (\nabla_{M}\partial_{N}\pi)^{2} - R^{MN} \partial_{M}\pi \partial_{N}\pi  = \frac{4}{9}m_6^2e^{-3\pi/2}R_{5},
\label{pieom1}
\end{eqnarray}
where $R_{MN}$ is the $5D$ Ricci tensor and $R_{5}$ is the Ricci scalar. Remarkably, even though the cubic $\pi$ interaction in~(\ref{5dcovcomplete}) has four derivatives, all higher-derivative terms cancel in the variation, yielding a second-order equation of motion for $\pi$.  This is a nontrivial and important property of the DGP $\pi$ lagrangian~\cite{Nicolis:2004qq}. In the decoupling limit of Fierz-Pauli massive gravity, by contrast, the $\pi$ lagrangian takes an analogous form, but its equation of motion is higher order --- there is a ghost mode propagating at the non-linear level~\cite{Boulware:1972zf,Deffayet:2005ys,Creminelli:2005qk,Dvali:2007kt}. See~\cite{Gabadadze:2009ja,deRham:2009rm} for an interesting recent proposal of a non-linear completion of Fierz-Pauli gravity that seemingly avoids these pitfalls.

Next we obtain the junction conditions at the brane position by setting to zero the boundary contributions to the variation of~(\ref{5dcovcomplete}). Assuming a $\mathbb{Z}_2$-symmetry, variation with respect to the metric yields the Israel junction condition
\begin{eqnarray}
\nonumber
	2M_{5}^{3} e^{-3\pi/2} \left( K q_{\mu\nu} - K_{\mu\nu} - \frac{3}{2} q_{\mu\nu} {\cal L}_{n} \pi \right) & = & \frac{27}{8}\frac{M_{5}^{3}}{m_{6}^{2}} \left( \partial_{\mu}\pi \partial_{\nu}\pi {\cal L}_{n} \pi + \frac{1}{3} q_{\mu\nu} \left( {\cal L}_{n}\pi \right)^{3} \right) \\
	& & + \ T^{(4)}_{\mu\nu} - M_4^2G_{\mu\nu}^{(4)} \,,
	\label{covjc1}
\end{eqnarray}
where
\be
	T_{\mu\nu}^{(4)} \equiv - \frac{2}{\sqrt{-q}} \frac{\delta (\sqrt{-q}{\cal L}_{\rm matter})}{\delta q^{\mu\nu}}
\ee
is the matter stress-energy tensor on the brane, and $G_{\mu\nu}^{(4)}$ is the Einstein tensor derived from the induced metric $q_{\mu\nu}$. Similarly, varying~(\ref{5dcovcomplete}) with respect to the scalar, we obtain after some algebra the boundary condition for $\pi$ on the brane:
\be
e^{-3\pi/2}K + \frac{9}{8m_{6}^{2}} \Big( K_{\mu\nu} \partial^{\mu}\pi \partial^{\nu}\pi + 2{\cal L}_{n}\pi \Box_{4}\pi + K ({\cal L}_{n} \pi)^{2} \Big) = 0\,.
\label{covjc2}
\ee
Equations~(\ref{covjc1}) and~(\ref{covjc2}) are not independent, of course; the divergence of~(\ref{covjc1}) can be shown to be proportional to~(\ref{covjc2}) after using the bulk momentum constraint equation. As a nontrivial check on our junction conditions, we have evaluated (\ref{covjc1}), (\ref{covjc2}) in a gauge in which the brane is at fixed position ($y=0$) and the bulk metric is time-dependent, and have shown that the result agrees with the boundary conditions obtained by integrating the bulk equations~(\ref{bulkein})--(\ref{pieom1}) across the delta-function sources at $y=0$ (see Appendix \ref{appendix}).


\section{The Cosmological Evolution on the Brane}
\label{cosmology}

The study of brane-world cosmology requires us to use our equations of motion to obtain a Friedmann equation on the brane, assuming homogeneity and isotropy along the 3+1 world-volume dimensions. The junction conditions~(\ref{covjc1}) and~(\ref{covjc2}) do not form a closed system of equations for $q_{\mu\nu}$, hence deriving an induced Friedmann equation requires knowledge of the bulk geometry~\cite{Chatillon:2006vw}. 

Because of the bulk scalar field, there is no Birkhoff's theorem to ensure that the bulk solutions are necessarily static under the assumption of homogeneity and isotropy on the brane --- the most general bulk geometry depends on both the extra-dimensional coordinate {\it and} time. For concreteness, however, we focus here on a static warped geometry with Poincar\'e-invariant slices,
\be
{\rm d}s^2_{\rm bulk} = a^2(y) (-{\rm d}\tau^2 + {\rm d}\vec{x}^2) + {\rm d}y^2\,.
\label{staticbulk}
\ee
While admittedly restrictive, we view this ansatz as a tractable first step in exploring cascading cosmology. And, as we will see, the resulting phenomenology is already surprisingly rich.

The brane motion is governed by two functions, $y(t)$ and $\tau(t)$, describing the embedding, where $t$ is proper time on the brane. The induced metric is of the Friedmann-Robertson-Walker (FRW) form, with spatially-flat ($k=0$) constant-time hypersurfaces,
\begin{eqnarray}
{\rm d}s^{2}_{\rm brane} = -{\rm d}t^{2} + a^{2}(y){\rm d}\vec{x}^2\,,
\label{4dmetric}
\end{eqnarray}
where, by virtue of $t$ being the proper time,
\begin{eqnarray}
	\left(\frac{{\rm d} t}{{\rm d}\tau}\right)^2 = a^2 - \left(\frac{{\rm d}y}{{\rm d}\tau}\right)^2\,.
\label{dtaudt}
\end{eqnarray}
Given a solution $a(y)$ to the bulk equations~(\ref{bulkein})--(\ref{pieom1}), the covariant junction conditions (\ref{covjc1}) and (\ref{covjc2})
allow us to solve for the embedding $(y(t),\tau(t))$, and hence the cosmology induced by brane motion through the warped bulk.

\subsection{A Dynamic brane in a static background}

With the static ansatz~(\ref{staticbulk}), the bulk equations~(\ref{bulkein})--(\ref{pieom1}) take on a form reminiscent of cosmological equations, with $a(y)$ acting as a scale factor as a function of ``time" $y$. In particular, the $(5,5)$ component yields a Friedmann-like equation
\be
\left(\frac{a'}{a}\right)^2 = \frac{a'}{a}\pi'\left(\frac{9}{8m_{6}^{2}} e^{3\pi/2} \pi'^{2} + 1\right)\,,
\label{bulkeq1}
\ee
whereas the $(\mu,\nu)$ components yield
\begin{eqnarray}
\frac{a''}{a} + \left(\frac{a'}{a}\right)^{2} = \frac{9}{16m_{6}^{2}} e^{3\pi/2} \pi'^{2}\pi'' -\frac{1}{2}\left( \frac{3}{2}\pi'^{2} - 3\frac{a'\pi'}{a} - \pi'' \right)\,.
\label{bulkeq2}
\end{eqnarray}
Meanwhile, the equation of motion for $\pi$ can be written as
\begin{eqnarray}
\frac{{\rm d}}{{\rm d}y}\left(\frac{a'}{a}\pi'^{\;2}\right)  + 4\left(\frac{a'\pi'}{a}\right)^{2} = - \frac{4}{9}m_6^2e^{-3\pi/2}\left[ 3\left(\frac{a'}{a}\right)^{2} + 2\frac{a''}{a} \right]  .
\label{bulkeq3}
\end{eqnarray}
As usual, the Bianchi identity guarantees that only two of these equations are independent. Finding exact solutions to these equations requires a numerical approach, which we will perform in section~\ref{numevol}. To offer analytical guidance, however, we seek approximate solutions to~(\ref{bulkeq1})--(\ref{bulkeq3}) in the so-called strong- (section~\ref{strong}) and weak-coupling (section~\ref{weak}) regimes in which the non-linear terms in $\pi$ respectively dominate or are negligible in these equations.

The brane embedding $(y(t),\tau(t))$ is determined by the junction conditions, which involve the extrinsic curvature tensor and the Lie derivative of $\pi$. Using~(\ref{dtaudt}) the relevant quantities are
\be
K^i_{\;j}  =  \frac{a'}{a} \sqrt{1 + \left(\frac{{\rm d}y}{{\rm d}t}\right)^{2}}\delta^i_{\,j}\;, \qquad K^0_{\; 0} =  \frac{1}{a} \frac{{\rm d}}{{\rm d}y} \left( a\sqrt{1 + \left(\frac{{\rm d}y}{{\rm d}t}\right)^{2}} \right)\,,
\ee
and
\be
{\cal L}_{n}\pi  =  \pi' \sqrt{1 + \left(\frac{{\rm d}y}{{\rm d}t}\right)^{2}}\,.
\ee
For the stress energy on the brane, we assume a collection of (non-interacting) perfect fluids with energy densities $\rho^{(i)}_{\rm m}$ and pressures $P^{(i)}_{\rm m}$, obeying the standard continuity equations
\begin{eqnarray}
	\frac{{\rm d}\rho^{(i)}_{\rm m}}{{\rm d}t} + 3H(\rho^{(i)}_{\rm m}+P^{(i)}_{\rm m}) = 0\,,
\label{staticfluid}
\end{eqnarray}
where $H\equiv {\rm d}\ln a/{\rm d}t$ is the Hubble parameter on the brane. These components may include baryonic matter, dark matter, radiation and a cosmological constant $\Lambda$. Equation~(\ref{staticfluid}) is consistent with the picture that matter is not allowed to flow into the bulk and is confined to the brane. 

It is clear, therefore, that given a bulk solution $a(y)$, a single junction condition is sufficient to solve for the cosmological evolution on the brane. Indeed, although (\ref{covjc1}) and (\ref{covjc2}) yield three equations, two of these follow from the bulk Hamiltonian and momentum constraints, which are automically satisfied given a solution $a(y)$. Since we are interested in the Friedmann equation on the brane, the natural choice is the $(0,0)$ component of~(\ref{covjc1}). Noting that $\partial_{0}\pi = \pi' {\rm d}y/{\rm d}t$ and ${\rm d}y/{\rm d}t = aH/a'$, we can write the resulting equation as the standard Friedmann equation with an additional effective energy density $\rho_{\pi}$ resulting from the $\pi$ field,
\begin{eqnarray}
3H^2M_4^2 &=& \sum_i \rho^{(i)}_{\rm m} + \rho_\pi\,,
\label{staticjc}
\eea
where 
\bea
\rho_{\pi}&\equiv &M_5^3 \sqrt{a'^{2} + a^{2}H^{2}}  \left\{\frac{9}{8m_6^2} \left( 2\left(\frac{aH}{a'}\right)^{2} - 1 \right) \frac{\pi'^{3}}{a'}-6 e^{-3\pi/2} \left( \frac{\pi'}{2a'} - \frac{1}{a} \right) \right\} \,,
\label{rhopi}
\eea
encoding all the complexity and new physics of our model. Given a solution $a(y)$, $\pi(y)$ to the bulk equations, we may invert this relation to obtain $y(a)$, and use this to express all $y$-dependent terms in $\rho_\pi$ as functions of $a$. Equation~(\ref{staticjc}), together with the continuity equations~(\ref{staticfluid}), then form a closed system for the brane scale factor $a(t)$. 

Before moving on to explicit solutions, we note in passing that $\rho_\pi$ is {\it not} positive definite. When combined with $\Lambda$, this can lead to an effective equation of state parameter $w < -1$ for the effective dark energy component. This phantom behavior already occurs in the normal branch of the standard DGP model~\cite{Lue:2004za,Sahni:2002dx,Chimento:2006ac}, a phenomenon that can be understood in the decoupling limit as arising from non-minimal coupling of the brane-bending mode to brane gravity~\cite{Chow:2009fm}. (It is well-known that $w<-1$ can be achieved in scalar-tensor theories when working in the Jordan frame~\cite{Boisseau:2000pr,Carroll:2004hc,Das:2005yj,Agarwal:2007wn}.) Similarly here, the scalar $\pi$ is kinetically-mixed with the brane graviton, which can lead to phantom behavior for dark energy.

\subsection{The Strong-coupling regime}
\label{strong}

By analogy with the Vainshtein screening mechanism around astrophysical sources, we expect that at early times, when the energy density in the universe is high, $\pi$ should be strongly coupled and cause small deviations from standard $4D$ Friedmann cosmology. In other words, the non-linear terms in $\pi$ dominate, but as a result $\rho_\pi$ is negligible compared to matter and radiation. Moreover, since the total variation in $\pi$ is expected to be small in this regime ($|\Delta \pi| \ll 1$), by rescaling $M_5$ we can assume that $e^{3\pi/2}\approx 1$.

Consider~(\ref{bulkeq1}) and (\ref{bulkeq2}) in the regime in which the non-linear terms in $\pi$ dominate:
\begin{eqnarray}
\nonumber
	\frac{a'}{a} & = & \frac{9}{8m_{6}^{2}} \pi'^{3}\; , \\
	\frac{a''}{a} + \left(\frac{a'}{a}\right)^{2} & = & \frac{9}{16m_{6}^{2}}\pi'^{2}\pi''\, .
\label{bulkeqstrong}
\end{eqnarray}
These admit scaling solutions, given by
\begin{eqnarray}
\nonumber
	a(y) & = & \left( \frac{12}{5}m_{6}|y| \right)^{5/12}, \\
	\pi(y) & = & \left(\sqrt{\frac{5}{4}}m_{6}|y|\right)^{2/3},
\label{strongbulk}
\end{eqnarray}
where we have a chosen a mass scale proportional to $m_{6}$ in the solution for $a(y)$. This leaves the scale factor today, $a_{0}$, to be a free parameter. It is straightforward to check that the above solution also satisfies the third bulk equation (\ref{bulkeq3}) in the strong-coupling approximation. The approximation $\pi\ll 1$ implicit in~(\ref{strongbulk}) is therefore valid provided $y\ll m_6^{-1}$. This defines the regime of validity of this solution.

The naked singularity at $y=0$ --- the analogue of a big bang singularity in cosmology --- introduces a plethora of complications if included as part of the bulk geometry. It is therefore safest to exclude this part of the geometry when performing the $\mathbb{Z}_2$ identification. As a result, however, the warp factor grows without bound as one moves away from the brane, which may indicate a strong-coupling problem. A related question concerns the stability of this solution --- by analogy, the self-accelerated branch of the DGP model also has a growing warp factor~\cite{Deffayet:2000uy} and is well-known to suffer from instabilities. We leave a careful study of these important issues to future work.

The above solutions for $\pi(y)$ and $a(y)$ can be used to express the effective Friedmann equation~(\ref{staticjc}) solely in terms of
the brane scale factor. In the strong-coupling regime, the $\pi'^{3}/m_{6}^{2}$ term dominates over the $\pi'$ term in~(\ref{rhopi}), giving
\begin{eqnarray}
	\rho_{\pi} \approx M_{5}^3 \left\{ 2 \frac{H^{2}}{m_{6}^{2}}a^{24/5} + 5 \right\} \sqrt{H^{2} + \left(\frac{m_{6}}{a^{12/5}}\right)^{2}} \,,
\end{eqnarray}
and~(\ref{staticjc}) reduces to
\be
H^2 \approx \frac{1}{3M_4^2}\sum_i \rho^{(i)}_{\rm m}  + m_5\left\{ \frac{2}{3} \frac{H^{2}}{m_{6}^{2}}a^{24/5} + \frac{5}{3} \right\} \sqrt{H^{2} + \left(\frac{m_{6}}{a^{12/5}}\right)^{2}}\,,
\label{friedstrong}
\ee
where $m_5$ is defined in~(\ref{m5}). Combined with the matter fluid equation~(\ref{staticfluid}), this effective Friedmann equation completely describes the evolution of the universe in the strong-coupling regime. 

In contrast with the standard DGP Friedmann equation, $H^2 = \rho/3M_4^2 \mp 2m_5H$, where the departure from $4D$ gravity is set by $H/m_5$, here the relative importance of $\rho_\pi$ also depends on a time-dependent scale $m_{6}/a^{12/5}$. In particular, for a fixed initial value of $a$, the magnitude of the modification can be set arbitrarily by a suitable choice of $m_{6}$. This freedom reflects the choice of initial condition for the brane motion in the bulk --- because the bulk is warped, different initial locations of the brane yield different expansion histories. In the standard DGP model, on the other hand, the bulk is flat Minkowski space, and hence all initial conditions (within the same branch of solutions) are related by the Poincar\'e group.

To proceed, we consider two limiting cases:

\begin{itemize}

\item If $H\gg m_{6}/a^{12/5}$, then the modification to the Friedmann equation further reduces to
\be
\rho_\pi \approx  2 M_{5}^3\frac{a^{24/5}H^{3}}{m_{6}^{2}} \,.
\label{rhopiv1}
\ee
Assuming that the universe is dominated by a matter component with general equation of state $w$, then $H \sim a^{-3(1+w)/2}$, and thus $\rho_\pi \sim a^{3(1-15w)/10}$. In terms of an effective equation of state for the $\pi$ field, defined through ${\rm d}\ln\rho_{\pi}/{\rm d}\ln a \equiv -3(1+w_\pi)$, we have
\be
w_{\pi} = -\frac{11}{10} + \frac{3}{2}w\,.
\ee
In particular, since $w_\pi < w$, it is clear that $\rho_\pi$ becomes more and more negligible as we look backwards in time. Moreover, in a universe dominated by baryonic and/or cold dark matter ($w=0$), the $\pi$ field can act as a dark energy fluid with phantom equation of state $w_{\pi} = -11/10$.

A phantom equation of state opens up the possibility of the $\pi$ field acting like dark energy and driving cosmic expansion.  In the strong regime, the Friedmann equation (\ref{staticjc}) can be approximated by the cubic equation,
\bea
AH^3 -3H^2 + \frac{\rho_m}{M_4^2} &=&0 \label{cubic}
\eea
with $\rho_\pi/M_4^2 = AH^3$, and $A= 2 a^{24/5}m_{5}/m_6^2$.
%
Differentiating (\ref{cubic}) gives
\bea
w_{\rm eff} \equiv-1 -\frac{2}{3}\frac{\dot{H}}{H^2}= -1-\frac{2}{3}\frac{1-\frac{13}{5}\Omega_\pi}{\Omega_\pi-\frac{2}{3}}.
\eea
For $\Omega_\pi > 5/24 \approx 0.21$, this gives $w_{\rm eff} < -1/3$, and acceleration occurs. However, the $\pi$ field is unable to dominate the energy density and fully account for the current phase of accelerated expansion, because of a singularity  at $\Omega_\pi=2/3$ for which $w_{\rm eff}\rightarrow -\infty$.

\item In the opposite regime, $H \ll m_{6}/a^{12/5}$, we have
\be
\rho_{\pi}\approx 5 M_{5}^3 \frac{m_{6}}{a^{12/5}} \,.
\ee
In this case, the $\pi$ component has a fixed effective equation of state, $w_{\pi} = -1/5$, independent of the matter on the brane. Again, this pushes the total equation of state to more negative values.

\end{itemize}

\subsection{The Weak-coupling regime}
\label{weak}

By analogy once again with the Vainshtein story in DGP, at late times we expect the non-linear terms in $\pi$ to be negligible, corresponding to gravity becoming higher-dimensional. In this approximation, the bulk equations (\ref{bulkeq1}) and (\ref{bulkeq2}) reduce to
\begin{eqnarray}
\nonumber
	\frac{a'}{a} & = & \pi'\;, \\
	\frac{a''}{a} + \left(\frac{a'}{a}\right)^{2} & = & -\frac{1}{2}\left( \frac{3}{2}\pi'^{2} - 3\frac{a'\pi'}{a} - \pi'' \right)\,,
\label{bulkeqweak}
\end{eqnarray}
which again admit a scaling solution
\begin{eqnarray}
\nonumber
	a(y) & = & \left(\frac{12}{5}m_{6}|y|\right)^{2/5}, \\
	\pi(y) & = & \frac{2}{5} \ln (m_{6}|y|)\,.
\end{eqnarray}
The mass scale in the solution for $a(y)$ has been chosen to be consistent with the strong-coupling solution. It is straightforward to check that this solution consistently satisfies the third bulk equation~(\ref{bulkeq3}) in the weak-coupling limit.

Substituting this solution into~(\ref{rhopi}), the effective energy density in $\pi$ in the weak-coupling regime reduces to
\begin{eqnarray}
\rho_{\pi} \approx 3M_{5}^3 \left\{ \left( \frac{12}{5} \right)^{3/5} \frac{1}{a^{3/2}} + \frac{3}{4} \frac{H^{2}}{m_{6}^{2}} \right\} \sqrt{H^{2} + \left(\frac{24}{25} \frac{m_{6}}{a^{5/2}}\right)^{2}}.
\label{friedweak}
\end{eqnarray}
In the limiting case in which $H \gg m_{6}/a^{5/2}$, this further reduces to
\begin{eqnarray}
	\rho_\pi \approx \frac{9}{4} \frac{M_{5}^3}{m_{6}^{2}} H^{3}\,,
	\label{rhopiv2}
\end{eqnarray}
which implies that
\begin{eqnarray}
	w_{\pi} = \frac{1}{2} + \frac{3}{2}w\,.
\end{eqnarray}
It is interesting to note that for a cosmological constant with $w = -1$, the $\pi$ field also behaves as a cosmological constant, $w_{\pi} = -1$. Similarly, for $H \ll m_{6}/a^{5/2}$,
\begin{eqnarray}
	\rho_\pi\approx \frac{72}{25} \left( \frac{12}{5} \right)^{3/5} \frac{M_{5}^3m_{6}}{a^{4}}\,,
\end{eqnarray}
which behaves like a relativistic component ($w_{\pi} = 1/3$) independent of the matter on the brane.

\subsection{Numerical Solutions}
\label{numevol}

\begin{figure}[!h]
  \begin{center}
    \includegraphics[width=4.5in,angle=0]{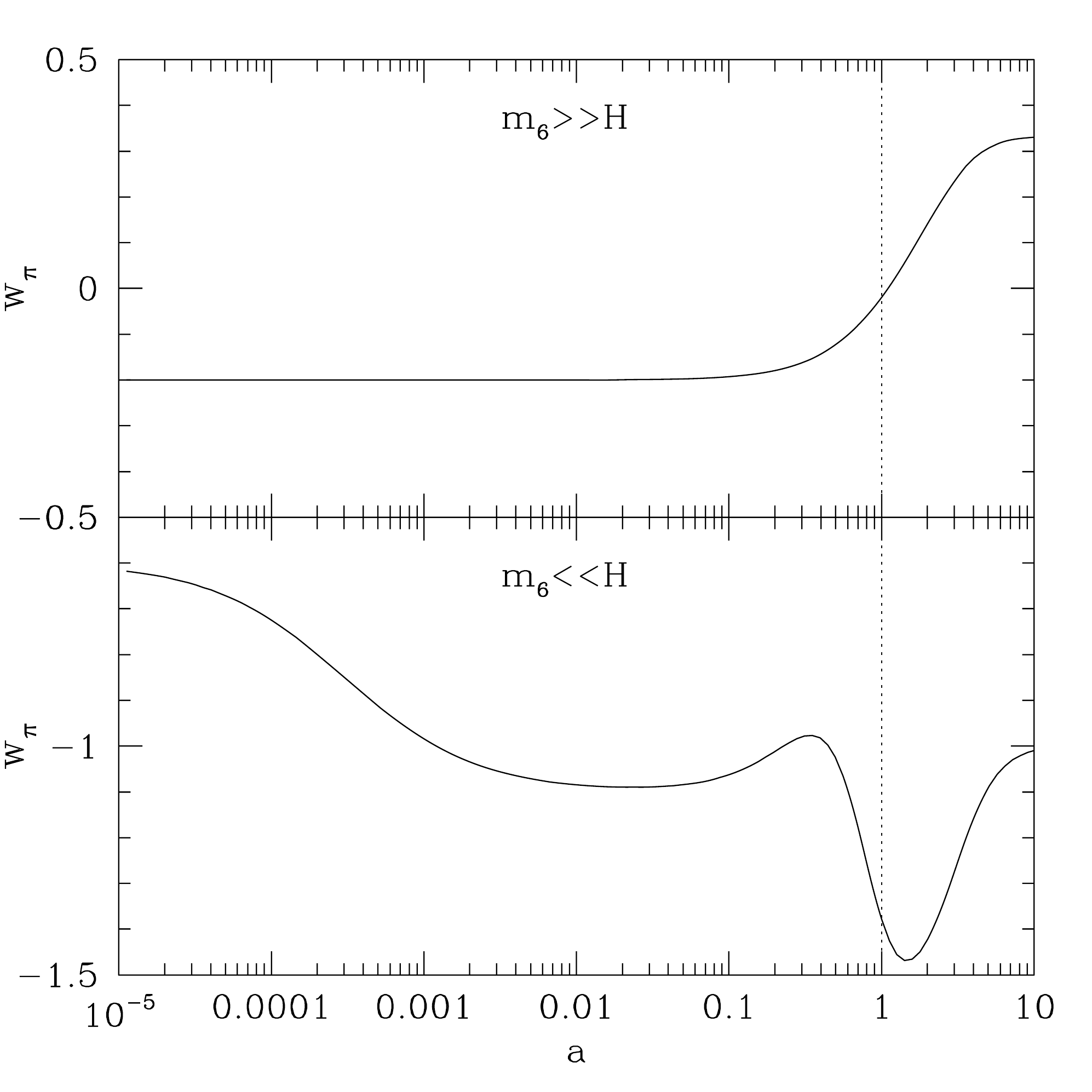}
    \caption{Evolution of the effective equation of state, $w_{\pi}=-1-(1/3){\rm d}\ln \rho_\pi/{\rm d}\ln a$, for the $\pi$-dependent modifications to the Friedmann equation (\ref{rhopi}).  The numerical results are consistent with the analytical predictions for the large (upper panel) and small (lower panel) $m_6$ limits in the strong- ($a\ll1$) and weak-coupling ($a\gg1)$ regimes. Here we use the numerical values (in natural units $c=\hbar=1$): $H_{0} = 2.33 \times 10^{-4}$ Mpc$^{-1}$ (i.e. $H_0=70$ km s$^{-1}$Mpc$^{-1}$), (upper panel) $m_{6} = 10^{30}$ Mpc$^{-1}$ ($m_{6} \gg H$)  and  $m_{5} = 10^{-40}$ Mpc$^{-1}$ and  (lower panel) $m_{6} = 10^{-15}$ Mpc$^{-1}$ ($m_{6} \ll H$) and  $m_{5} = 10^{-30}$ Mpc$^{-1}$. The $\pi$ field is a subdominant component of the total energy density at all times, and late-time acceleration is driven by $\Lambda$.} \label{fig1}
  \end{center}
\end{figure}

\begin{figure}[!t]
  \begin{center}
    \includegraphics[width=4.5in,angle=0]{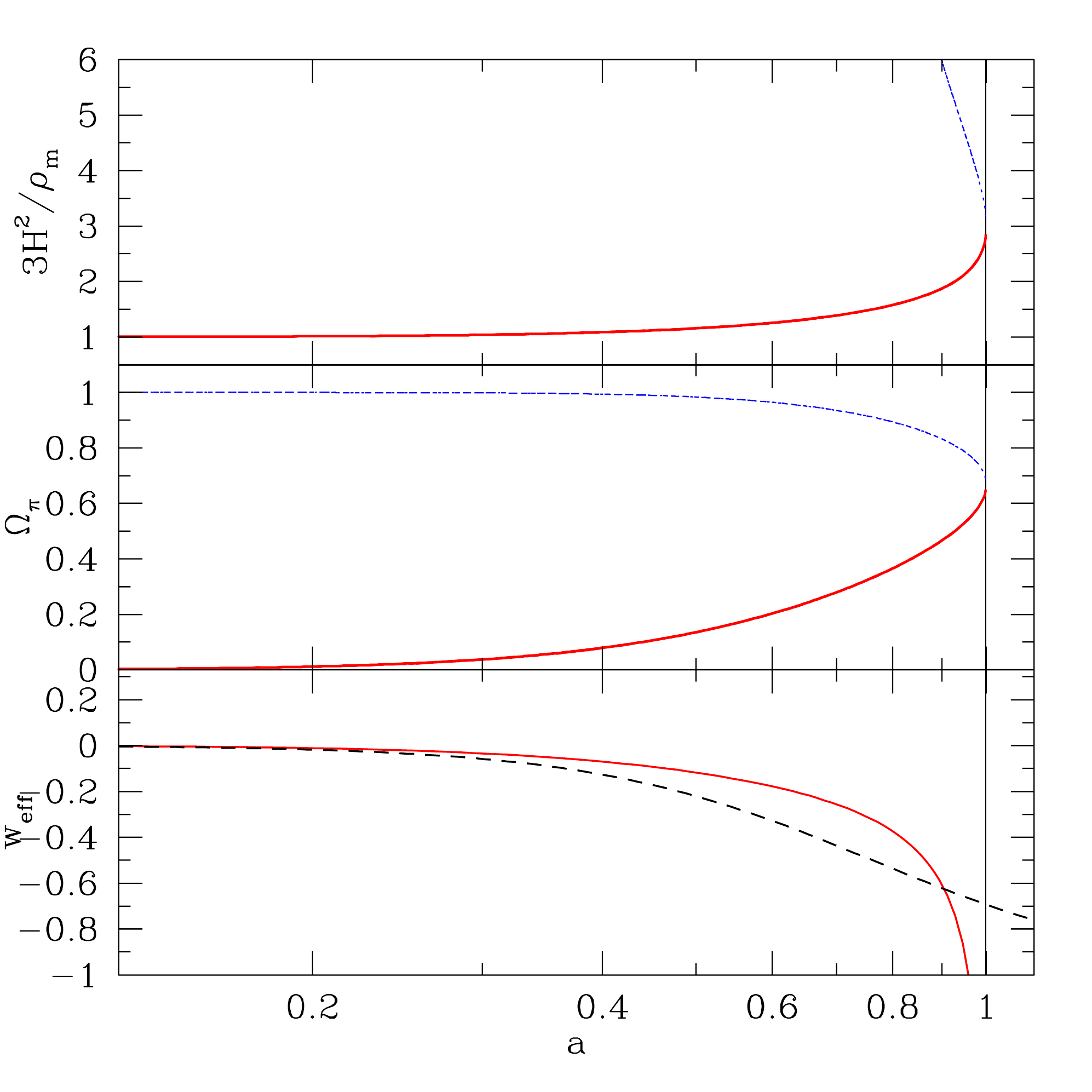}
    \caption{Example evolution histories in which no cosmological constant is present to drive cosmic acceleration. [Top panel] The deviation of the expansion history from that derived from standard matter (for which $3H^2/\rho_m=1$). The blue and red curves each show consistent solutions to the modified Friedmann equation (\ref{staticjc}): one solution (red, thick line) recovers the standard expansion history at early times and then undergoes accelerated expansion at late times; the other solution (blue, dotted line) has an expansion history entirely inconsistent with that of standard $\Lambda$CDM, with the $\pi$ field dominating the expansion at all eras, and undergoing heavily decelerated expansion at late times. [Center panel]  
The evolution of the effective fractional energy density, $\Omega_{\pi}=8\pi G\rho_\pi/3H^2$, for the two solutions discussed above. For the accelerating solution, the phantom-like behavior in the matter era  allows the $\pi$ field to dominate and drive cosmic acceleration at late times. The model is not physical, however, since as $\Omega_\pi\rightarrow 2/3$ one finds $\dot{H}\rightarrow \infty$ and a singularity occurs. [Bottom panel] A comparison of the effective equation of state for the expansion, $w_{\rm eff} = -1-(2/3){\rm d}\ln H/{\rm d}\ln a$, for the accelerating $\pi$ (red, full line) and fiducial $\Lambda$CDM (black, dashed line) scenarios. For the $\pi$ driven expansion histories, we use the numerical values $H_{0} = 2.33 \times 10^{-4}$ Mpc$^{-1}$, $m_{6} = 3.5\times 10^{-18}$ Mpc$^{-1}$ and $m_{5} = 4.4\times 10^{-31} $Mpc$^{-1}$ for which the maximum singularity occurs just after $a=1$.} \label{fig2}
  \end{center}
\end{figure}

To complement the analytical strong and weak-field limits  in \ref{strong} and \ref{weak}, we numerically evolve the full bulk and brane equations given in (\ref{bulkeq1})--(\ref{bulkeq3}) and (\ref{staticfluid})--(\ref{rhopi}), in the presence of matter on the brane. We assume zero spatial curvature on the brane, and include relativistic and pressureless components consistent with the standard cosmological model: $\Omega_{\rm m}=0.3$, $\Omega_{\rm r} =8.5\times 10^{-4}$. We further fix the scale factor today to be $a_0=1$. 

Starting well into the radiation dominated era, with $a\ll1$, we evolve $\pi$, $\pi'$, $y$ and $t$ forward with respect to $\ln a$:~(\ref{bulkeq2}) and~(\ref{bulkeq3}) combine to form an equation for $\pi''$, from which we form an equation for ${\rm d} \pi'/{\rm d}\ln a= \pi'' /(a'/a)$, and (\ref{staticjc}) can be rewritten as a cubic equation in $H^2$, which, if a positive, real solution exists, can be used to evolve $t$ through ${\rm d}t/{\rm d}\ln a = 1/H$. 

In Fig.~\ref{fig1},  we show numerical confirmation of the analytical dynamical attractor solutions for $w_{\pi}$ discussed in \ref{strong} and \ref{weak}. For scenarios with $m_6\ll H$ we find attractor solutions of $w_\pi=-0.6$ and $-1.1$ in the (strongly-coupled) radiation and matter dominated eras respectively, and $-1$ in the (weakly-coupled) $\Lambda$-dominated epoch. For $m_6\gg H$, strongly- and weakly-coupled attractors arise with $w_\pi=-0.2$ and $w_\pi=1/3$, respectively. 

The $\pi$ field has an effective `phantom' equation of state in the matter-dominated ($m_6\ll H$) regime. This opens up the apparent possibility of cosmic acceleration arising within cascading cosmology without the need for a cosmological constant. However, as discussed in Sec.~\ref{strong}, while it is possible to generate acceleration at late times, one hits a singularity in the expansion history when $\Omega_\pi = 8\pi G \rho_\pi/3H^2=2/3$ so that the universe cannot smoothly transition towards $\Omega_\pi\rightarrow 1$. In Fig.~\ref{fig2} we show a realization of such a scenario, with the onset of cosmic acceleration, and the limiting presence of the singularity.

This singularity is of an unusual nature --- it is not equivalent to the Big Rip scenarios in which $H$ and $a$ both become infinite in a finite space of time, since the Hubble parameter $H$ and scale factor $a$ remain finite while $\dot{H}$ diverges. Moreover, the bulk geometry is smooth at that point, and it is the brane embedding that is singular. It is possible that this singularity could be circumvented by the use of a more general metric ansatz than the static case considered here to obtain solutions on the brane, or by accounting for finite brane-thickness effects. We leave this to future investigation.
 

\section{Conclusions}
\label{conclusions}

Cascading gravity is a phenomenologically rich framework for exploring new phenomena associated with infrared-modified gravity, and offers a promising avenue for realizing degravitation. This construction circumvents many of the technical hurdles of earlier attempts at higher-dimensional extensions of DGP: the induced propagator is free of divergences, the theory is perturbatively ghost-free, and adding a small tension on the $4D$ brane yields a bulk solution which is nowhere singular and remains perturbative everywhere. Due to its higher-dimensional nature, however, extracting cosmological predictions presents a daunting challenge.

In this paper we have considered the more tractable problem of a $5D$ effective brane-world set-up, obtained from the full $6D$ cascading theory through the decoupling limit. Strictly speaking, the decoupling limit leaves us with an action describing a scalar $\pi$ and weak-field gravity, which is therefore of limited use for studying cosmology. But since $\pi$ is responsible for most of the interesting phenomenology of cascading gravity, we have proposed a fully covariant, non-linear $5D$ completion of the decoupling theory, as a proxy for the complete $6D$ model. Our effective action describes $5D$ DGP gravity with a bulk $\pi$ scalar field, coupled to a $4D$ brane with intrinsic gravity. 

Upon supplementing the $5D$ action with boundary terms (to yield a well-posed action principle), we obtained covariant junction conditions across the brane, relating the extrinsic curvature to delta-function sources on the brane. In order to study cosmology on the brane, we then considered a scenario in which a dynamic brane moves across a static bulk, and consistently solved the bulk and brane equations of motion. We derived analytical solutions for the induced cosmology in the strong- and weak-coupling regimes, valid at early- and late-times, respectively, and confirmed these expectations with a complete numerical analysis. 

Thanks to a cosmological Vainshtein mechanism, the bulk scalar $\pi$ and the helicity-0 mode of the $4D$ massive graviton both decouple at early times, resulting in an early-universe cosmology that closely reproduces the expansion history of the standard big bang theory. At late times, however, these scalar modes effectively contribute to dark energy through a modification of the Friedmann equation and result in small deviations from $\Lambda$CDM expansion at late times. Although these scalars thus affect dark energy, a singularity in the brane embedding prevents the modification from being entirely responsible for cosmic acceleration.

We are currently studying the evolution of cosmological perturbations in this context. Such an analysis should also shed light on the all-important question of stability. With our branch choice, the modification to the Friedmann equation behaves as an effective component with {\it positive} energy density. At first sight this is worrisome, since the counterpart in standard DGP is the self-accelerated branch, which is plagued with ghost instabilities. It is crucial to investigate whether or not this is the case here too. From a phenomenological perspective, we are performing a full likelihood analysis for the predictions of the model, including both expansion and growth histories. On much smaller scales, we are also working to derive the consequences for Lunar Laser Ranging observations, thereby generalizing the analysis of~\cite{Dvali:2002vf} for DGP to the degravitation/cascading framework~\cite{Dvali:2007kt}.


\section*{Acknowledgements}
We thank Claudia de Rham, Kurt Hinterbichler and Andrew Tolley for helpful discussions. N.A., R.B. and M.T. are supported by NASA ATP grant NNX08AH27G. N.A. and R.B.'s research is also supported by NSF CAREER grant AST0844825, NSF grant PHY0555216 and by Research Corporation. The work of J.K. is supported in part by funds provided by the University of Pennsylvania. M.T. is also supported by NSF grant PHY-0930521, and by Department of Energy grant DE-FG05-95ER40893-A020.

\appendix
\section{Junction conditions for a specific choice of metric}
\label{appendix}

As a non-trivial check on our covariant junction conditions, in this Appendix we show that they agree with direct integration of the bulk equations of motion, when specialized to a gauge in which the brane is at fixed coordinate position. Varying the action (\ref{5dcovcomplete}) with respect to the metric and $\pi$ led to the 5D Einstein field equations (\ref{bulkein}) and the $\pi$ equation of motion (\ref{pieom1}). Upon adding in contributions from delta-function sources on the brane to the bulk Einstein field equations we obtain,
\begin{eqnarray}
	e^{-3\pi/2} G_{MN} & = & -\frac{27}{16m_6^2} \left[ \partial_{(M}(\partial\pi)^{2}\partial_{N)}\pi - \frac{1}{2}g_{MN}\partial^{K}(\partial\pi)^{2}\partial_{K}\pi - \partial_{M}\pi\partial_{N}\pi\Box_{5}\pi \right] \nonumber \\
	& & - \left( g_{MN}\Box_{5} - \nabla_{M}\nabla_{N} \right) e^{-3\pi/2} + \frac{\delta(y)}{b} {\delta_{M}}^{\mu} {\delta_{N}}^{\nu} M_{5}^{-3} \left( T_{\mu\nu}^{(4)} - M_{4}^{2} G_{\mu\nu}^{(4)} \right),
\label{bulkeinapp}
\end{eqnarray}
where $T_{\mu\nu}^{(4)}$ and $G_{\mu\nu}^{(4)}$ are defined in section \ref{eoms}. The $\pi$ equation is as before,
\begin{eqnarray}
	(\Box_{5}\pi)^{2} - (\nabla_{M}\partial_{N}\pi)^{2} - R^{MN} \partial_{M}\pi \partial_{N}\pi  = \frac{4}{9}m_6^2e^{-3\pi/2}R_{5}.
\label{pieomapp}
\end{eqnarray}
We are interested in studying brane-world cosmological solutions, for which we specialize to $5D$ spacetime metrics of the form,
\begin{eqnarray}
	{\rm d}s^{2}_{\rm bulk} & = & -n^{2}(\tau,y){\rm d}\tau^{2} + a^{2}(\tau,y){\rm d}\vec{x}^{2} + b^{2}(\tau,y){\rm d}y^{2},
\label{metric}
\end{eqnarray}
with $\pi = \pi(\tau,y)$. The brane is defined by the hypersurface $y = 0$, and we compute the junction conditions, relating the jump across the brane of $a'$, $n'$, and $\pi'$ or the extrinsic curvature to delta-function sources on the brane, for this metric. 

The metric is required to be continuous across the brane in order to have a well-defined geometry. However, its derivatives with respect to $y$ may be discontinuous across $y=0$, and therefore the second derivatives with respect to $y$ will contain a Dirac delta function \cite{Binetruy:1999ut}, so that $a'' = \widehat{a''} + [a']\delta(y)$, and similarly for $n$. Here $\widehat{a''}$ is the standard derivative (the non-distributional part of the double derivative of $a$), and $[a']$ is the jump in the first derivative across $y=0$. We similarly allow $\pi$ to be discontinuous across the brane and write, $\pi'' = \widehat{\pi''} + [\pi']\delta(y)$. We further impose a $\mathbb{Z}_2$-symmetry ($y \leftrightarrow -y$) across the brane, and therefore $[X'] \equiv 2X'(0^{+})$ for the metric and $\pi$ discontinuities.

In order to obtain the junction conditions we integrate the (0,0) and $(i,j)$ components of the bulk field equations (\ref{bulkeinapp}) and the $\pi$ equation of motion (\ref{pieomapp}) over an infinitesimal region of the extra dimension $y$, spanning $y=0$. This picks out coefficients of $\delta(y)$ and leads to the following three junction conditions,
\begin{eqnarray}
	e^{-3\pi_{0}/2} \left( \frac{a'_{0}}{a_{0}b_{0}} - \frac{1}{2}\frac{\pi'_{0}}{b_{0}} \right) - \frac{9}{16}\frac{1}{m_{6}^{2}} \left( \frac{1}{3} \frac{\pi'^{2}_{0}}{b_{0}^{2}} - \frac{\dot{\pi}^{2}_{0}}{n_{0}^{2}} \right) \frac{\pi'_{0}}{b_{0}} = -\frac{1}{6M_{5}^{3}} \sum_{i} \rho^{(i)}_{\rm m} + \frac{1}{2m_{5}} \frac{\dot{a}^{2}_{0}}{a^{2}_{0}n^{2}_{0}},
\label{ajunction}
\end{eqnarray}
\begin{eqnarray}
	e^{-3\pi_{0}/2} \left( \frac{n'_{0}}{n_{0}b_{0}} - \frac{1}{2}\frac{\pi'_{0}}{b_{0}} \right) &-& \frac{9}{16}\frac{1}{m_{6}^{2}} \left( \frac{1}{3} \frac{\pi'^{2}_{0}}{b_{0}^{2}} + 2\frac{\dot{\pi}^{2}_{0}}{n_{0}^{2}} \right) \frac{\pi'_{0}}{b_{0}} \nonumber \\
	& =&  \frac{1}{6M_{5}^{3}}\left(3\sum_{i} P^{(i)}_{\rm m} + 2\sum_{i} \rho^{(i)}_{\rm m}\right) - \frac{1}{2m_{5}n_{0}^{2}} \left( \frac{\dot{a}^{2}_{0}}{a^{2}_{0}} + 2 \frac{\dot{a}_{0}\dot{n}_{0}}{a_{0}n_{0}} - 2\frac{\ddot{a}_{0}}{a_{0}} \right), \nonumber \\
\label{njunction}
\end{eqnarray}
\begin{eqnarray}
	e^{-3\pi_{0}/2} \left( 3\frac{a'_{0}}{a_{0}} + \frac{n'_{0}}{n_{0}} \right) + \frac{9}{4}\frac{1}{m_{6}^{2}} \pi'_{0} \left\{ -\frac{\ddot{\pi}_{0}}{n_{0}^{2}} + \frac{\dot{\pi}_{0}}{n_{0}^{2}} \left( \frac{\dot{n}_{0}}{n_{0}} - 3\frac{\dot{a}_{0}}{a_{0}} \right) \right\} \nonumber \\
	+ \ \frac{9}{8}\frac{1}{m_{6}^{2}} \left\{ \frac{n'_{0}}{n_{0}} \left( \frac{\pi'^{2}_{0}}{b_{0}^{2}} - \frac{\dot{\pi}_{0}^{2}}{n_{0}^{2}} \right) + 3\frac{a'_{0}}{a_{0}}\frac{\pi'^{2}_{0}}{b^{2}_{0}} \right\} = 0,
\label{pijunction}
\end{eqnarray}
where the subscript 0 indicates that the function is evaluated at $y=0$. 

We have checked that (\ref{ajunction})--(\ref{pijunction}) agree exactly with the covariant junction conditions, (\ref{covjc1}) and (\ref{covjc2}), specialized to this gauge. Further, they reduce to those of the standard DGP model \cite{Deffayet:2000uy} in the strong coupling limit $\pi \rightarrow 0$, $m_{6} \rightarrow 0$, in which the bulk scalar $\pi$ decouples.


\bibliographystyle{apsrev}
\bibliography{degravitation}

\end{document}